\documentstyle[11pt,newpasp,twoside,epsf]{article}
\def\edcomment#1{\iffalse\marginpar{\raggedright\sl#1\/}\else\relax\fi}
\reversemarginpar

\begin{document}
\title{A birth and growth of a collimated molecular jet from an AGB star}
\author{Hiroshi Imai}
\affil{Joint Institute for VLBI in Europe, Postbus 2, 7990 AA Dwingeloo, 
the Netherlands}
\affil{Department of Physics, Faculty of Science, Kagoshima University, 
1-21-35 Korimoto, Kagoshima 890-0065, Japan}
\author{Philip~J.~Diamond}
\affil{Jodrell Bank Observatory, University of Manchester,
Macclesfield, Cheshire SK11 9DL, UK}

\begin{abstract}
With the VLBA, we have observed water masers associated with 
the OH/IR star, W43A, which trace a birth and growth of a molecular jet. 
The water masers exhibit the collimated distribution (1700~AU: 20~AU) and 
fast motions ($\pm$150~km~s$^{-1}$). The maser distribution was well fit 
by a precessing jet model. The jet length has extended by 800~AU within 
8~years, indicating that the extension rate is roughly equivalent to the jet 
speed (150~km~s$^{-1}$). Very likely the jet was born around the year 1960. 
An elongated planetary nebula will be formed by such a jet during an extremely 
short period ($<<$ 1000 yrs) of the transition to form a ptoto-planetary nebula. 
Then fading of the water maser jet will be observed during our own life. 
\end{abstract}

\section{Introduction}

Evolved stars of about a solar mass generally are spherically symmetric, yet 
the planetary nebulae (PNe) that immediately follow them on their evolutionary 
path often are not. Collimated stellar jets of material have been observed up to 
0.3~pc from central objects of planetary nebulae, and precessing jets have been 
proposed as the origin of the asymmetries in the subsequent planetary nebulae 
(e.g., Sahai \& Trauger 1998). Moreover, it has recently been shown theoretically 
that magnetic fields could launch and shape the jets (Blackman~et~al.\ 2001). 

We consider here a member of a small, but exceptionally interesting class of 
stellar water maser sources that are characterized by extremely large spreads 
of maser velocities (up to 260~km~s$^{-1}$, Likkel, Morris, \& Maddalena 1992). 
This class of "water fountains" includes only three {\it bona fide} objects: 
IRAS~16342$-$3814 (Morris, Sahai, \&  Claussen 2003), IRAS~19134$+$2131 
(Imai~et~al.\ 2003) and W43A (Diamond \& Lyman 1988; Imai~et~al.\ 2002). 
This class of sources appear to be closely related morphologically and kinematically 
to other post-AGB stars revealed by optical, infrared and mm-wave observations. 

Here we report VLBA observations of the W43A water masers during 
1994--2002, which present a birth and growth of a stellar jet with precession, 
occurring formed prior to forming a proto-planetary nebula. 

\section{VLBA observations}

The VLBA observations have been made in the following epochs, 1994 June 25 
and October 10, 1995 March 17, 2002 April 3, July 26, and November 24. We 
observed W43A and calibrators for 10~hrs in total in each of the epochs. Since 
the year 2002, the phase-referencing technique has been applied to precisely 
estimate the source coordinate with respect to the stable reference frame fixed 
with extragalactic quasars and the W43A 1612-MHz OH masers. 

\begin{figure}
\epsfxsize=12cm
\plotone{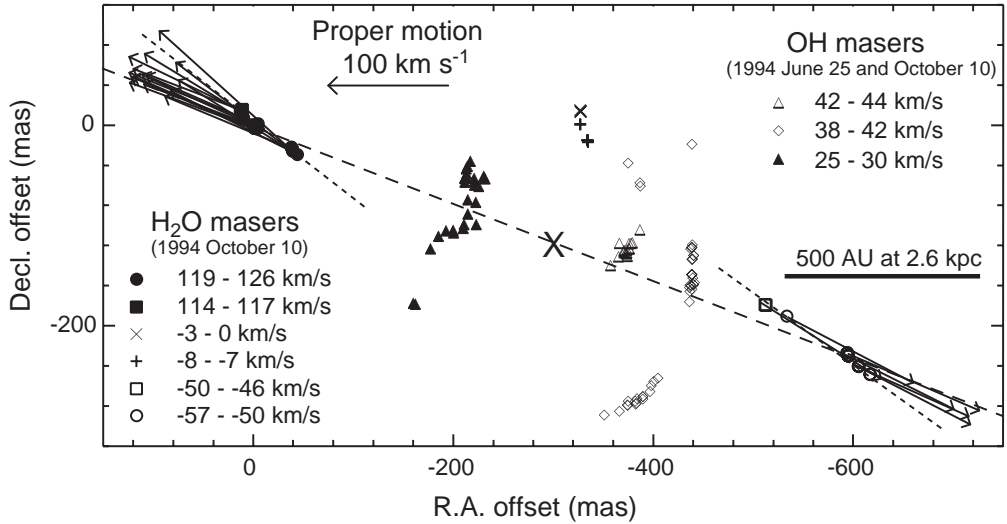}
\caption{Spatial distributions of H$_{2}$O masers found on 1994 October 10 
and 1612-MHz OH masers found on 1994 May 26 and October 10 in W43A 
(Imai~et~al.\ 2002). Although the relative positions between the H$_{2}$O 
and OH masers are under investigation, they are roughly expected to be 
those in the figure. Dotted lines show spatial alignments of maser features 
in the red-shifted and blue-shifted H$_{2}$O maser clusters. A broken line 
shows the direction of the jet. The OH masers are well modeled by a 
spherically-expanding flow with an expansion velocity of $\sim$9~km~s$^{-1}$ 
and a radius of $\sim$500 AU.}
\end{figure}

\section{Growing and precessing stellar jet}

The W43A jet traced by water maser emission is highly collimated, both 
spatially and kinematically (see figure 1). Most of the maser spots are 
concentrated in the receding (north--east side) and approaching 
(south--west side) clusters. Both clusters have lengths of 250--350 AU but 
widths of only $\approx$20~AU. The two clusters are separated by 
$\approx$1700~AU. The width of the aligned masers is only $\sim$1/85 
of the total length. The ratio is much smaller than any of those observed in
"molecular outflows". Measured proper motions with line-of-sight velocities 
of the masers indicate the presence of a bipolar jet with a 3-D space speed of 
145~$\pm$~20~km~s$^{-1}$. Moreover, two pairs of the maser clusters are 
located almost point-symmetrically with respect to the estimated dynamical 
center of the water masers, which is the most likely origin of the jet 
(Imai~et~al.\ 2002). The extremely high collimation and point-symmetrical 
distributions of similar jets have been seen in optical observations of PNe and 
young stellar objects, but this is the first such molecular jet with these 
characteristics to be observed. 

Furthermore, we found that the observed angular pattern of the water masers 
has grown during 1994--2002, the growth rate is roughly same as the speed of 
the maser feature motions. The angular pattern is well fit by a precessing jet 
model (figure 2). These results support that the jet was born around the year 
1960 and precessing since its birth. 

\begin{figure}
\epsfxsize=12cm
\plotone{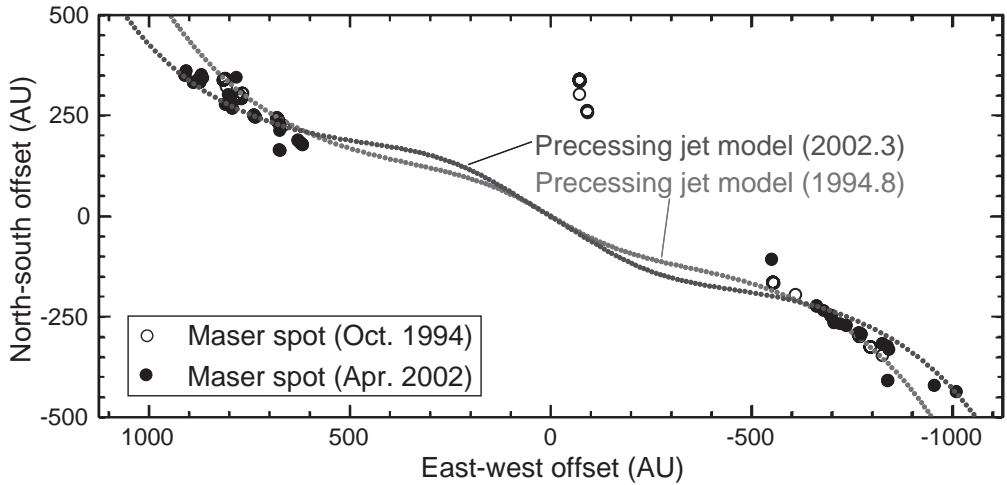}
\caption{The angular distribution of the H$_{2}$O masers detected on 1994  
October 10 and 2003 April 3, fitted by the spiral patterns expected from a 
precessing jet model. The modelled jet has a constant velocity of 
150~km~s$^{-1}$ and a jet axis with an inclination of 39\deg\ with respect 
to the sky plane, a position angle of 65\deg, and an axis precession with an 
angular amplitude of 5\deg\ and a period of 55 years. We assumed that only 
the direction of the bipolar ejection of material is varying continuously with 
time. The dynamical center of the jet has a systemic radial velocity of 
34~km~s$^{-1}$ and a relative position of ($\Delta\alpha =$ $-$296 mas, 
$\Delta\delta=$ $-$112 mas) in Figure~1.}
\end{figure}

\section{Water fountain before becoming a planetary nebula}

Previous observations have inferred that stellar jets have already been 
formed $\sim$1000~yrs before an AGB star evolves into a post-AGB star 
and starts the photoionization of its circumstellar envelope
(Miranda~et~al.\ 2001). On the other hand, the dynamical age of the jet  
in W43A is only $\approx$ 40~yrs. Combined with the presence of SiO 
masers in W43A (Nakashima \& Deguchi 2003), which are detected in very 
few PPNe and with the OH masers such as those typically found in OH/IR 
stars (figure 1), this estimate strongly suggests that we are observing a star 
in transition. Another two "water fountains" also have similar dynamical 
ages ($\sim$100~yrs for IRAS~16342$-$3814, Morris, Sahai, \&  Claussen 2003, 
50~yrs for IRAS~19134$+$2131, Imai~et~al.\ 2003), but their morphologies 
seem already to be destroyed; it is difficult to find clear alignment and point 
symmetry in their maser distributions. Very likely, the water fountains last 
for a very short period ($\leq$100~yrs). The latter two sources have already 
exhibited (bipolar) nebulosity (Sahai~et~al.\ 1999; Sahai~et~al.\ 2003 in 
preparation), while W43A does not exhibit any nebulosity (Deguchi 2003 in 
private communication). Thus the water fountain appears just before or 
during the transition to a PPN. Comparing the above time scales with the 
duration of significant mass-loss rate from OH/IR stars (1000-4000~yrs, 
Lewis 2001), a water fountain should affect the development of PN 
morphology from the beginning of its formation (Sahai \& Trauger 1998). 

\section{Future prospective}

The driving object and the formation mechanism of the W43A jet are still 
unclear. W43A could be an AGB star able to create a collimated jet
driven by the magnetic force due to the dynamo action at the interface
between the rapidly rotating core and the more slowly rotating
envelope of the star (Blackman~et~al.\ 2001)  Alternatively, W43A could 
be a binary system where the ejected material from a mass-losing star falls 
onto an accretion disk surrounding a companion that creates the jet. These 
hypothesis are examined by astrometric technique for the water and hydroxyl 
masers in W43A, they are associated with the collimated jet and a 
spherically-expanding envelope, respectively. The locations of the dynamical 
centers of these jet and envelope should provide a strong constraint on 
binary system models. Therefore, we are performing phase-referencing 
VLBI observations to determine their coordinate with respect to the common 
extragalactic quasars. The location of the SiO masers also precisely indicate 
the position of the star that create the envelope and provides material to the 
jet. 

Furthermore, because the devolution of the W43A jet is expected before our 
death, we will monitor the water maser morphology and kinematics for elucidating 
a growth of a planetary nebula developing its asymmetrical morphology. 

\acknowledgments
The NRAO's VLBA is a facility of the National Science Foundation, 
operated under a cooperative agreement by Associated Universities, Inc. 
H.~I.\ was partially financially supported by the Leids Kerkhoven-Bosscha 
Fonds of Leiden University for attending the conference.


\begin{references}
\reference
Blackman,~E.~G., Frank,~A., Markiel,~J.~A., Thomas,~J.~H., \&
Van Horn,~H.~M.\ 2001, Nature, 409, 485
\reference
Diamond, P.J., \&  Nyman, L.-A.\ 1988, in IAU Symp.\ 129, 
VLBI and their impact on Astrophysics, ed. J.M.~Moran \&  M.J.~Reid 
(Dordrecht: Reidel), 249
\reference
Imai,~H., Morris,~M., Sahai,~R., Hachisuka,~K., \& 
Azzollini,~J.~R.~F.\ 2003, \aap, submitted
\reference 
Imai,~H., Obara,~K., Diamond,~P.~J., Omodaka,~T., \&  Sasao,~T.\ 2002, 
Nature, 417, 829
\reference
Lewis,~B.~M.\ 2001, \aap, 560, 400
\reference
Likkel,~L., Morris,~M., \&  Maddalena,~R.~J., 1992, \aap, 256, 581 
\reference
Miranda,~L.~F., G\'omez,~Y., Anglada,~G., \& Torrelles,~J.~M.\ 
2001, Nature, 414
\reference
Morris~M.~R., Sahai,~R., \&  Claussen,~M.\  2003, RmxAC, 15, 20
\reference
Nakashima,~J., Deguchi,~S., 2003, \pasj, 55, 229
\reference
Sahai,~R., te~Lintel~Hellert,~P., Morris,~M., Zijlstra,~A., \&  Likkel,~L.\ 
1999, \apj, 514, L115 
\reference
Sahai,~R., \&  Trauger,~J.~T.\ 1998, AJ, 116, 1357
\end{references}
\end{document}